# Scintillation characteristics of undoped and $Cu^+$-doped $Li_2B_4O_7$ single crystals

Masaaki Kobayashi[1,*], Mitsuru Ishii[2], Nachimuthu Senguttuvan[2,#]

[1]*IPNS, High Energy Accelerator Research Organization(KEK),
Oho, Tsukuba, 305-0801 Japan*
[2]*Shonan Institute of Technology (SIT), Fujisawa, 251-8511 Japan,*

Scintillation characteristics of undoped and $Cu^+$-doped lithium tetraborate $Li_2B_4O_7$ (LTB) were studied including optical transmittance, photo-luminescence, radioluminescence for X- and γ-rays, α/γ ratio, and decay kinetics. The total time-integrated *LY*s in undoped and $Cu^+$-doped LTB for X-rays are ~600 and ~760 ph/MeV (photons/MeV), respectively. The decay kinetics in undoped and $Cu^+$-doped LTB are similar to each other. Typical decay spectra for pulsed X-rays can be fitted with four exponentials: for fast ($\tau_1$~0.8 ns, $\tau_2$~25-50 ns), medium ($\tau_3$~300-400 ns), and slow ($\tau_4$~20-30 μs) components. The slow component occupies about 60% of the total *LY*, while the fast ones less than 10%. The 10-90% rise time was 163 ps. The α/γ ratio was 0.18 for external $^{241}Am$ α-rays. The obtained increase in *LY* due to $Cu^+$ doping remains modest. The $Cu^+$-induced emission contains both fast and slow components, requiring further studies of the emission mechanism to explain the fast component.

## INTRODUCTION

While X-ray analysis is a powerful tool in studying the structure of heavy (large *Z*) materials, its power decreases for light materials. As biotechnology progresses in recent years, reactions of thermal neutrons have attracted strong interest because they give important information on light organic materials including the molecular structure as well as the dynamics of DNA as an example. In the study of organic molecules, scattering and diffraction of neutrons are important tools to determine the positions of light atoms such as H, C and O. Thermal neutrons ($n_{th}$) are mostly used for the neutron beam due to the obtainable high intensity, less ambiguity in the energy and large capture cross sections in optimized detectors.

-------------------

* Corresponding author. *E-mail address*: masaaki.kobayashi@kek.jp
# Present address: New Business Development Headquarters, Hitachi Chemical Co. Ltd., Tokyo 100-6606, Japan.

At present, $^6$LiF/ZnS:Ag and $^6$Li-glass:Ce$^{3+}$ are widely used for efficient $n_{th}$ detectors [1]. While $^6$LiF/ZnS:Ag has a merit of large $LY$ of 175000 ph/$n_{th}$ (photons/$n_{th}$), the exponential decay constant τ is as large as 250 ns (Table 1). On the contrary, $^6$Li-glass:Ce$^{3+}$ has a smaller τ of ~60 ns due to the Ce$^{3+}$ emission center, while the $LY$ is at a reasonable level of 5000 ph/$n_{th}$. Both of them have two problems for the neutron detectors of the next generation: first, τ is large and second, the sensitivity to background γ-rays is not small enough due to heavy elements contained in the host matrices.

Table 1 Typical scintillators for $n_{th}$ detectors; ρ=density, $X_0$=radiation length, $\lambda_{em}$=emission peak wavelength, $\lambda_{aten}$ =attenuation length for $n_{th}$.

| Scintillator | ρ g/cm$^3$ | $X_0$ cm | $\lambda_{em}$ nm | $LY$(n) ph/$n_{th}$ | $LY$(γ) ph/MeV | α/γ ratio | decay τ ns | hygroscopic | $\lambda_{aten}$ mm | Notes and Ref. |
|---|---|---|---|---|---|---|---|---|---|---|
| $^6$Li-glass :Ce$^{3+†}$ | ~2.5 | 10.9 | 395 | 5000 | 4000 | 0.23 | 60 | no | 0.676 | $^6$Li-95% [2-4] |
| $^6$LiF/ZnS :Ag$^¥$ | 2.36 | ~6.0 | 450 | 175000 | 18000 | 2 | 250 | no | ~0.99 | $^6$Li-95% [2] |
| $^6$LiI:Eu$^{2+}$ | 4.06 | 2.18 | 470 | 53000 | 13000 | 0.19 | 1400 | strong | 0.613 | $^6$Li-95% [5,6] |
| LiBaF$_3$ :Ce$^{3+}$ | 5.27 | 2.08 | 220 310 325 | 2600$^\#$ | 2600$^\#$ | 0.22 | 1 15000 30 | no | 9.0 | natural Li [7,8] |
| Li$_2$B$_4$O$_7$ :Cu$^+$ | 2.42 | 16.4 | 305, 365 | ~56$^\#$ | 310$^\#$ | 0.18 | <1,30, 300, 25000 | no | 0.362 | natural Li,B [&] |
| B-loaded PS$^\$$ | 1.026 | ~40 | 425 | 1900 | 8000 | ~0.1 | 2.2 | no | 4.67 | natural B [2] |

†) GS20; the composition being 56 wt% SiO$_2$-18Al$_2$O$_3$-4MgO-18Li$_2$O-4Ce$_2$O$_3$.
¥) BC-704,   $) plastic scintillator BC-454.
#) Slow components with τ much larger than 1 μs are removed for practical use.
[&]=present work.

Lithium tetraborate doped with Cu$^+$, Li$_2$B$_4$O$_7$:Cu$^+$ or LTB:Cu$^+$, has a faster emission and may be one of the better candidates for the neutron detectors. It contains both $^{10}$B and $^6$Li in the host, being highly sensitive to $n_{th}$. The $e^{-1}$ attenuation length $\lambda_{aten}$ of LTB is ~0.362 mm for natural Li and B. This is even smaller than that of $^6$LiF/ZnS:Ag (~0.99 mm) and $^6$Li-glass:Ce$^{3+}$ (0.676 mm) for 95% enriched $^6$Li as seen in Table 1. This means that LTB may be much cheaper than $^6$LiF/ZnS:Ag and $^6$Li-glass:Ce$^{3+}$. LTB has a



relatively small density of 2.42 g/cm$^3$ and a small effective atomic number $Z_{eff}$ (defined for photoelectric effect dominance for γ-rays) of 7.33 due to absence of large-$Z$ elements in the host, keeping the sensitivity to γ-ray background rather small. Its radiation length and Moliere radius are 16.4 cm and 4.63 cm, respectively. The technique to grow large single crystals of LTB has been established for the application in TLD radiation monitors as well as surface acoustic waves (SAW) devices. The advantages mentioned above may justify the study of scintillation characteristics of single crystals of undoped and Cu$^+$-doped LTB even if the *LY* may be modest.

LTB is congruently [9] crystallized into the tetragonal system at a melting point of 917 $^o$C [10]. It is rugged and nonhygroscopic. LTB has a band gap of ~9.0 eV and a fundamental absorption edge of 7.3 eV (170 nm) [11-13]. The top of the valence band and the bottom of the conduction band are mainly occupied by O and B, respectively[11-13]. According to the study of undoped LTB with SOR [14,15], the intrinsic emission consists of two bands peaked around 3.3 eV (375 nm) and 4.2 eV (295 nm); the former has a faster decay with much smaller intensity than the latter. Both emission bands can be excited at ~8.0 eV (155 nm) peak close to the fundamental absorption edge. According to the hypothesis [14,15] for the luminescence mechanism, excitation close to the fundamental absorption edge creates excitons, and they should be partly relaxed to self-trapped excitons (STE) which should experience radiative recombination. The 3.3 and 4.2 eV emission bands should correspond to the σ and π components, respectively, of the STE luminescence. The decay consists of fast components (τ<1 ns, and ~8.5 ns) and a slow one (τ~µs range): the slow component is dominant both at room temperature and at 9.6 K. The *LY* does not change much by cooling the crystal from the room temperature down to 80 K. Thermal quenching sets in above 320 K [14,15].

Effects of Cu$^+$ doping on the scintillation characteristics were extensively studied including the emission spectra and decay kinetics for UV excitation [16-18], dependence of optical transmittance and photoluminescence on Cu$_2$O concentration in the melt [9,19], and emission spectra, *LY*, α/γ ratio and decay kinetics for radioluminescence [20]. According to the detailed study of LTB:Cu$^+$ for UV excitation [18], the emission band at ~365 nm is enhanced for two excitation peaks at ~240 nm (5.2 eV) and ~175nm (7.1 eV). This band may be either a new band or corresponds to the 375 nm band seen in undoped LTB. If the former hypothesis (new band) is taken, a natural explanation for the two excitation peaks will be in terms of parity forbidden



$3d^{10} \rightarrow 3d^9 4s^1$ and parity-allowed $3d^{10} \rightarrow 3d^9 4p^1$ transitions of the ground state of $Cu^+(3d^{10})$. As $3d^9 4s^1$ should be lower in energy than $3d^9 4p^1$, the 365 nm emission band should arise from the $3d^9 4s^1 \rightarrow 3d^{10}$ transition. Since the radioluminescence peak occurs at the same wavelength as for photoluminescence, it should be natural to take it again due to the $3d^9 4s^1 \rightarrow 3d^{10}$ transition of $Cu^+$. The slow decay with a single component of $\tau \sim 25$ µs [18] supports the hypothesis of parity forbidden $3d^9 4s^1 \rightarrow 3d^{10}$ radiative transition for this emission peak. There is no evidence for energy transfer from the host to $Cu^+$ [18]. There are still critical viewpoints for the $Cu^+$ emission mechanism mentioned above. Besides the hypothesis of the intracenter transitions in $Cu^+$ sketched above, the transitions between $Cu^+$ and ligand ($O^{2-}$) due to the MLCT or LMCT charge transfers [21] may also be possible. For an example, the excited electron and hole may be trapped in Cu and $O^-$, respectively, and the self-trapped electron and the self-trapped hole may experience radiative recombination. Although such emission due to recombination between trapped electron at defects and trapped hole may be possible without $Cu^+$-doping, $Cu^+$ doping may increase this type of emission intensity. The existence of slow decay components also in undoped LTB supports the hypothesis as sketched above. More studies on decay kinetics are necessary in both undoped and $Cu^+$-doped LTB.

With an interest in possible new fast-response detectors of $n_{th}$ and a hope to obtain experimental data which may help understanding the radio-luminescence mechanism of LTB:$Cu^+$, we evaluated in the present work the scintillation characteristics of single crystals of undoped and $Cu^+$-doped LTB including photoluminescence, radioluminescence due to X- and γ-rays, α/γ ratio, *LY*, rise time and decay kinetics, and radiation damage due to γ-rays.

## MEASURING METHODS

Single crystals of LTB were grown with a vertical Bridgmann technique using a platinum crucible [9,19]. Reducing atmosphere of nitrogen was employed in the growth of Cu-doped crystals while ordinary air for undoped ones. Test samples of $10 \times 10 \times 20$ mm$^3$ were cut from grown ingots of a typical size of 90 mm in length and 20 mm in diameter. The crystals were transparent and colorless except for the case of heavy Cu-doping which caused segregation of the dopant to some extent. The test samples were polished to the optical grade on the whole surface or on the four of six planes, depending on the necessity in measurements. Cu doping was done by adding



$Cu_2O$ powder to the melt. The $Cu_2O$ concentration was in a range of 0.0025 to 0.05 wt% (weight %). The $Cu_2O$ concentration by $x_w$ wt% corresponds to $Li_{2-x}Cu_xB_4O_7$ with $x=0.0236\ x_w$.

Transmission was measured with a spectrophotometer (Hitachi U2310). Excitation-emission spectra were obtained with a fluoro-spectrophotometer (Hitachi F4500). Phosphorescence spectra were also measured with the F4500 by starting 2 ms after termination of excitation.

Radioluminescence (RL) was measured using an X-ray generator (50 kV, 30 mA, Cu target) and a spectrometer PMA-11 (Hamamatsu, Photonic Multichannel Analyzer). The PMA-11 consists of a grating monochromator, an image intensifier and a linear CCD array. The obtained RL spectrum was corrected for the quantum efficiency of the photo-sensor in the spectrometer. Emitted light was transmitted to the spectrometer via quartz fiber from the injected surface of X-rays at 90 degrees so that the sample thickness (transmission) should not be important. *LY* was estimated by comparing the RL intensity of the LTB crystals with a single crystal of BGO ($Bi_4Ge_3O_{12}$, 10x10x30 mm$^3$, polished all over to the optical grade). The integration over the wavelength was done offline so as to cover the emission peaks.

Energy spectra of γ- and α-ray sources were measured by viewing the 1 cm thick crystal with a 2-inch PMT on each end (Fig. 1). Silicone grease OKEN6262 was applied between the crystal and the PMT windows, while the rest of the crystal surface was wrapped with aluminium foil. Coincidence of both PMT signals was used to reduce noises. The pulse height of the PMT on one end (Hamamatsu R2256-02, bialkali, silica window) was actually analyzed with an ADC (LeCroy QVT) in the charge mode within 1.1 μs gate, and multiplied by a factor of two in the result presented in this paper. A $^{60}$Co γ-ray and a $^{241}$Am α-ray sources were used. The α-ray source was attached onto one of the side surfaces of the crystal. The average injection frequency of α-rays to the crystal was 10 Hz.

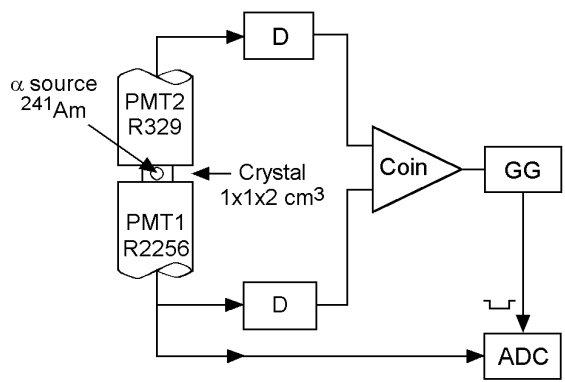

Fig. 1
Setup for *LY* measurement for α- and γ-rays. For measuring γ-rays, a $^{60}$Co source was placed at a short distance to the crystal. D: discriminator, GG: gate generator.



Rise and decay spectra were measured with a conventional single photoelectron technique using a pulsed X-ray source (width ~ 40 ps FWHM, repetition 100 Hz) [22]. A light-excited X-ray tube N5084 (Hamamatsu) with a tungsten anode, multialkali photocathode S20 and a thin Be window for X-rays was employed at a high voltage of 20 kV between photocathode and anode. This is essentially a single-stage PMT. The excitation light pulse was generated in a laser diode using InGaN in PiLAS (Picosecond Injection Laser, Advanced Laser-diode Systems, Berlin) and transported to the X-ray tube through a quartz fiber. The generated X-rays were collimated and hit the crystal surface placed at 45 degrees to the X-ray axis. The scintillation light was detected by a 2-inch PMT (Hamamatsu R2256-02, bialkali, silica window). The light intensity reaching the PMT was reduced to an order of magnitude smaller than one photoelectron per pulse so as to avoid distortion of the obtained decay spectrum. The PMT output signal was fed to a discriminator and its timing was registered in a CAMAC TDC with respect to the initial clock pulse which triggered the laser-diode.

Radiation damage was studied by irradiating the samples with $^{60}$Co γ-rays at Japan Atomic Energy Research Institute (JAERI). Three cycles of irradiation followed by transmission measurement were carried out to cover the accumulated dose from $10^3$ to $10^5$ Gy by increasing the accumulated dose by a factor of ten per cycle. The irradiation periods were 1h, 1h, and 22h for $10^3$, $10^4$, and $10^5$ Gy, respectively. The transmission measurements were carried out in 23h, 31h, and 22h after irradiation by $10^3$, $10^4$, and $10^5$ Gy, respectively. To compare the radiation damage in samples having different thickness, we calculated the irradiation-induced absorption coefficient $μ_{ir}$ defined as

$$μ_{ir} = (1/d)ln(T_0/T), \tag{1}$$

where $d$ is the sample thickness across which the transmittance $T_0$ (before irradiation) and $T$ (after irradiation) were measured.

## RESULTS

**Transmittance spectra**

Typical transmittance spectra of undoped and Cu-doped LTB are compared in Fig. 2. The undoped LTB shows no absorption dip in the wavelength region 200-600 nm of the present measurement. The Cu-doped LTB shows a dip at ~240 and 265 nm [23] corresponding to the broad excitation peak for the emission peak at 365 nm (Fig. 3).

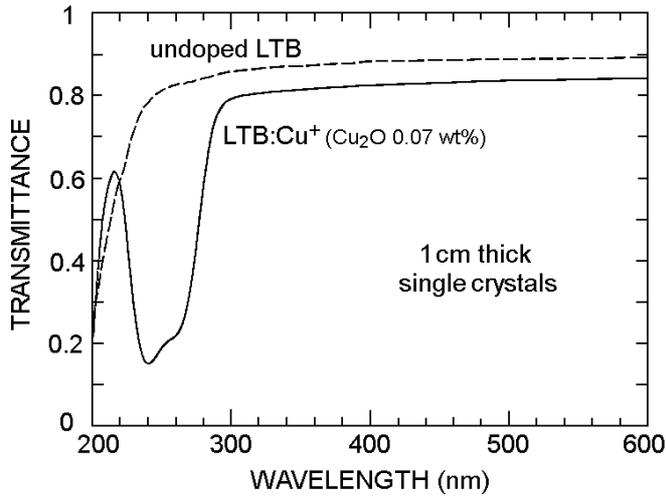 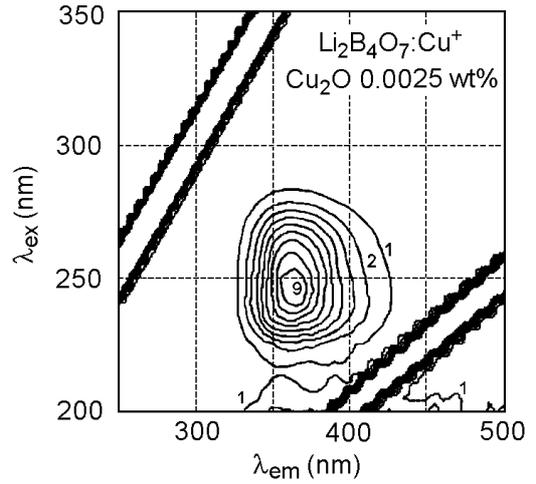

Fig. 2 Transmittance spectra of undoped and Cu$^+$ doped LTB single crystals across 1 cm thickness.

Fig. 3 Excitation-emission spectrum of Li$_2$B$_4$O$_7$:Cu$^+$. The numbers attached to the contour give the relative intensity

**Photoluminescence**

Figure 3 gives an excitation-emission spectrum from the surface of a single crystal of Cu-doped LTB. The contour of the emission intensity is plotted on the ($\lambda_{ex}$, $\lambda_{em}$) plane, where $\lambda_{ex}$ and $\lambda_{em}$ denote excitation and emission wavelengths, respectively. An emission peak sits at $\lambda_{em} \sim$ 365 nm for the excitation at $\lambda_{ex} \sim$ 240 nm in consistency with the radioluminescence spectrum shown later. A weak phosphorescence peak with a time constant less than 0.2 ms (the sensitivity limit of F4500) was also seen for $\lambda_{em} \sim$ 365 nm and $\lambda_{ex} \sim$ 240 nm. This is consistent with the slow component in radioluminescence to be described later. For undoped LTB, no sizable emission was excited for the excitation wavelength larger than 200 nm, the lower limit available in the F4500 spectrometer. This is consistent with the band gap of 9.0 eV mentioned in Introduction.

**Radioluminescence for X-rays**

Radioluminescence spectra of undoped and Cu-doped LTB for X-rays are compared in Fig. 4 with BGO. The emission peak of undoped LTB sits at ~305 nm. When Cu$^+$ is doped, an additional peak occurs at ~365 nm. The time-integrated *LY* of the emission peaks can be obtained by comparing the area of the peaks (220-600 nm) with the BGO (250-700 nm). Taking the *LY* of 1 cm thick BGO as ~6000 ph/MeV [24], the *LY* of undoped LTB and LTB:Cu$_2$O 0.07 wt% of Fig. 4 are 600 ph/MeV (10% of BGO) and 760 ph/MeV (12.7%), respectively. While the *LY* of the 305 nm peak remains roughly the same with plus minus 10% for samples of different $Cu_2O$ concentrations, that of the 365 nm peak scatters depending on both the $Cu_2O$ concentration and the crystal quality. From the segregation of Cu observed in crystal growth, the scattering of the *LY* of the 365 nm peak must be due to the scattering of the $Cu^+$ concentration and its uniformity in the crystal.

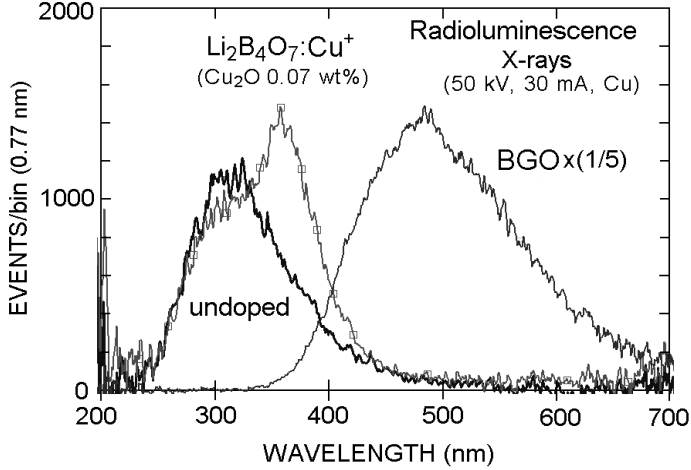

Fig. 4 Radioluminescence spectra of undoped and Cu-doped LTB single crystals in comparison with $Bi_4Ge_3O_{12}$ multiplied by 1/5.

**Measurement of α- and γ-rays**

Figure 5 gives energy spectra of 5.49 MeV α-rays from an $^{241}$Am source in undoped and Cu-doped LTB. When a $^{60}$Co γ-ray source was placed close to the crystal in the same measuring setup for α-rays, only a Compton edge, expected at 1.038 MeV, was seen as shown in Fig. 6. No total absorption peak was detected as LTB contains no large-*Z* atoms. While the Compton peak is usually broadened due to light yield statistics, a numerical simulation [25] shows that its position approximately corresponds to three fourths of the height of the broadened peak as indicated in Fig. 6. The above measurements gave the following number of photoelectrons (phe) for α- and γ-rays;

$N_{phe}$~ 3.2 phe/MeV for α and ~21.2 phe/MeV for γ-rays in undoped LTB  (2)

$N_{phe}$~ 4.2 phe/MeV for α and ~23.8 phe/MeV for γ-rays in LTB:$Cu^+$    (3).

From the comparison between α- and γ-rays, we obtained the α/γ ratio of ~0.15 in undoped LTB and ~0.18 in LTB:$Cu^+$. This is not far from the values (~0.2) in many inorganic scintillators, and smaller than 0.3 given in [20].

The *LY*s registered in the γ-ray spectra within 1 μs were estimated by comparing LTB and BGO crystals with the same setup for $^{60}$Co γ-ray source. As the decay constant of BGO is ~300 ns, almost all (~95%) of the total *LY* (6000 ph/MeV) lies in the γ-ray spectrum. Comparing the ADC position of the Compton edge (1.038 MeV) in the LTB spectrum with the total absorption peak (1.25 MeV on the average of 1.13 and 1.37 MeV) in BGO, we found 4.62% and 5.18% for the *LY*s of undoped LTB and LTB:$Cu^+$, respectively (Table2). They are smaller than the time-integrated *LY*s obtained for dc X-rays. The difference should be due to much slower component than 1 μs. Then the slow/total *LY* ratio should be 54% and 59% in undoped LTB and LTB:$Cu^+$, respectively.

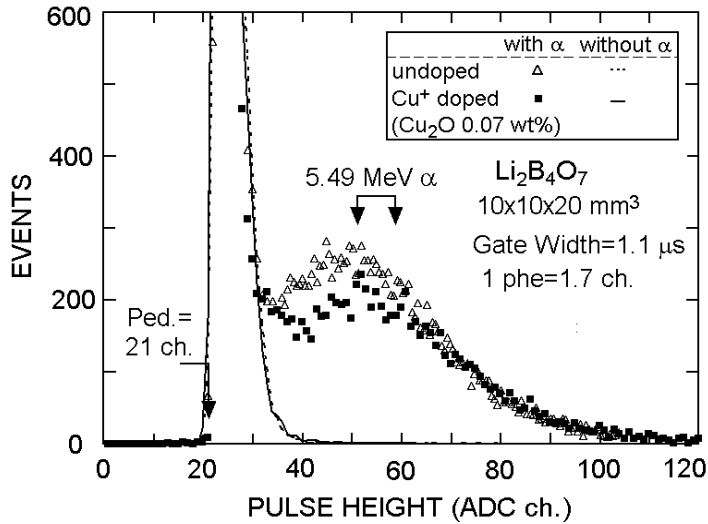

Fig. 5
Energy spectra of 5.49 MeV α-rays of $^{241}$Am in LTB:$Cu^+$ (solid squares) and undoped LTB (open triangles). The background spectra without the α-source are given with a solid (or dotted) curve for LTB:$Cu^+$ (undoped LTB).

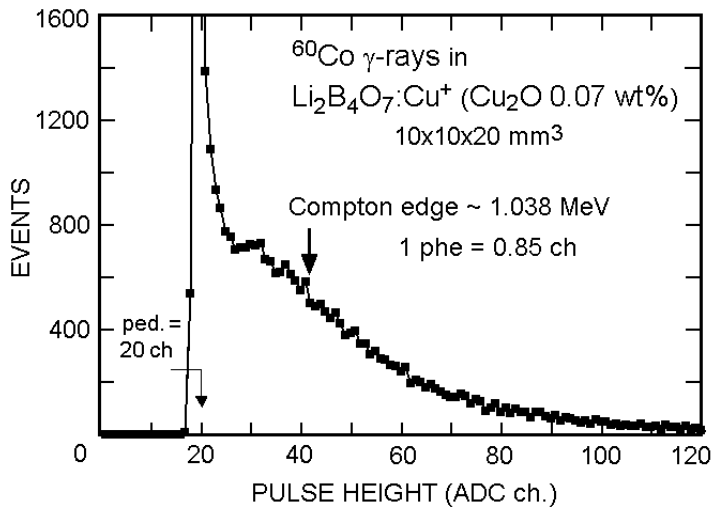

Fig. 6
Energy spectra of $^{60}$Co γ-rays in LTB:$Cu^+$.

Table 2   *LY* ratio of LTB to BGO.   *LY* in ph/MeV in parentheses are derived by taking 6000 ph/MeV for 1 cm thick BGO [24].

|  | undoped LTB | LTB:$Cu^+$ ($Cu_2O$ 0.07 wt%) |
|---|---|---|
| for X-rays (dc) | 10% (600 ph/MeV) | 12.7% (760 ph/MeV) |
| for pulsed γ-rays (1 μs) | 4.62% (277 ph/MeV) | 5.18% (311 ph/MeV) |
| slow/total *LY* ratio | 54±5%[*] | 59±5%[*] |

*) ±5% comes from typical ambiguity in the Compton peak position.

**Decay kinetics**

The 10-90% rise time was less than 0.2 ns for both undoped and $Cu^+$-doped LTB as shown in Fig. 7. It includes the *TTS* (transit time spread) of the PMT (R4998), whose nominal value is 160 ps FWHM. If the rise is approximated in a Gaussian shape and corrected for the *TTS* in quadrature, we obtain the net 10-90% rise time less than 163 ps.

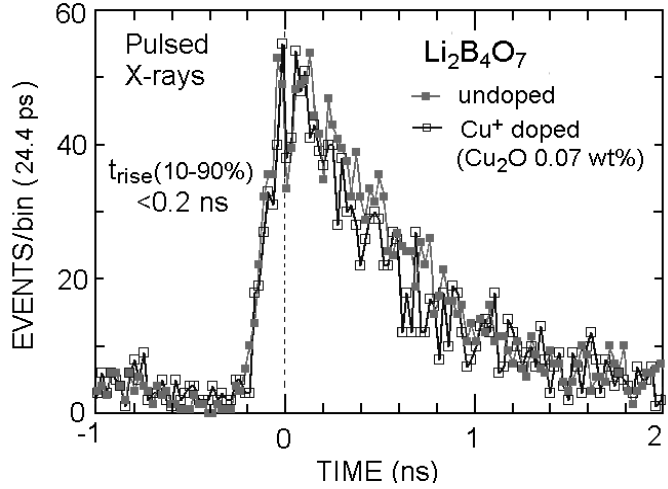

Fig. 7 Rise time of undoped and $Cu^+$-doped LTB. Decay spectra were registered for pulsed X-rays with a TDC of 4096 ch for 100 ns full scale.

Decay spectra of undoped and $Cu^+$-doped LTB are given in Fig. 8(a) and (b), respectively, for the full range of 400 ns. They can be fitted with 3 exponentials plus a constant. The fastest component has a decay constant $\tau_1 \sim 0.8$ ns, the second fast one $\tau_2 \sim 30$ ns and the third one $\tau_3 \sim 300$ ns for both undoped and $Cu^+$-doped LTB (Table 3).

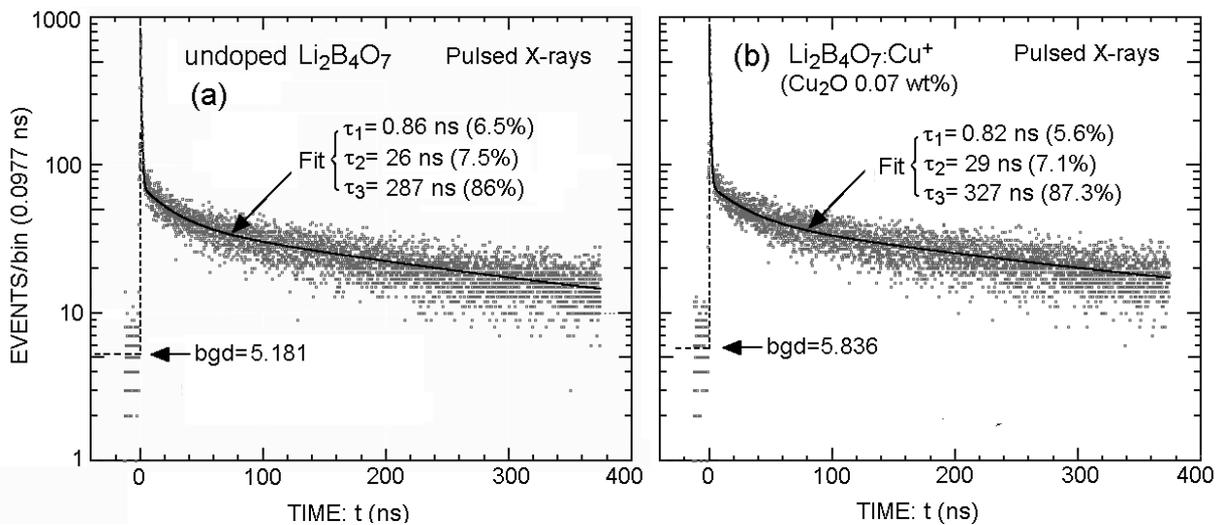

Fig. 8 Decay spectra of (a) undoped LTB and (b) LTB:$Cu^+$ for 400 ns full scale. In the fit of each spectrum, the constant background was fixed at the level of the average intensity for $t<-0.4$ ns.

Table 3 Decay characteristics for pulsed X-ray excitation in 0~400 ns range. Decay constants $\tau_1$ - $\tau_3$ are given with the percentage of *LY* in parentheses.

|  | undoped LTB | LTB:Cu$^+$ (Cu$_2$O 0.07 wt%) |
|---|---|---|
| 1st component | $\tau_1$=0.86 ns (*LY*~6.5%) | $\tau_1$=0.82 ns (*LY*~5.6%) |
| 2nd component | $\tau_2$=26 ns (*LY*~7.5%) | $\tau_2$=29 ns (*LY*~7.1%) |
| 3rd component | $\tau_3$=287 ns (*LY*~86.0%) | $\tau_3$=327 ns (*LY*~87.3%) |

While the dominant part of the *LY* can be fitted with $\tau_3$~300 ns component within the 0-400 ns range, a slower component with $\tau$ much larger than 1 μs is necessary to explain the difference of *LY*s between γ-ray (pulse) and X-ray (dc) excitations as described before. The existence of such slower component is also suggested in the decay spectra of Fig. 8 by a large difference of the tail of spectra from the constant background level. To study the slow component, we measured the decay spectra in 0-5 μs range. Typical decay spectra obtained are given in Fig. 9(a) and (b) for undoped and Cu$^+$-doped LTB, respectively. Since the TDC resolution is limited to 5 μs/4096ch=1.221 ns/ch, the subnanosecond $\tau_1$ component was not included in the fit. Then, each decay spectrum was fitted for t>4.9 ns with a constant plus 3 exponentials. The constant term was fixed at the level of the average intensity for t<−4.9 ns.

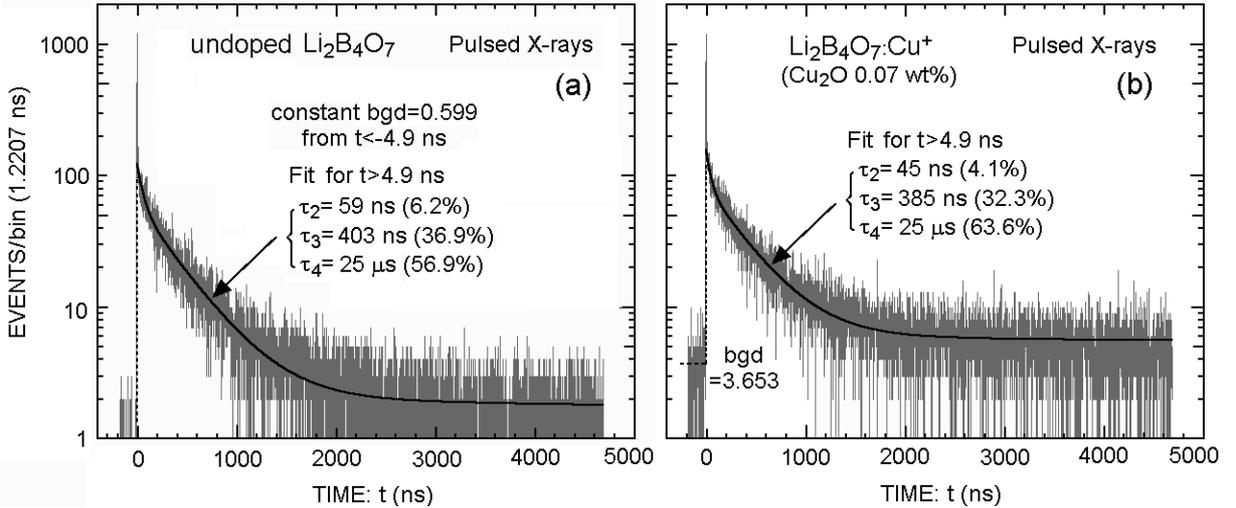

Fig. 9 Decay spectra of (a) undoped LTB and (b) LTB:Cu$^+$ for 5000 ns full scale. Each spectrum was fitted for t>4.9 ns by omitting the subnanosecond $\tau_1$ component and fixing the decay constant $\tau_4$ of slow component at 25 μs (see the text). The constant background was fixed at the the average intensity for t<−4.9 ns.

In the fit in Fig. 9, the decay constant $\tau_4$ of the slowest component was fixed at 25 μs due to the following reason. Actually $\tau_4$ could not be determined precisely only from the fit, because reasonable fit was obtained even if $\tau_4$ was changed largely above ~ 1 μs. Significant change in $\tau_4$ can be compensated in the fit by a small change in $\tau_3$ because the full range of 5 μs is not large enough. Then, we restricted $\tau_4$ from the slow/total *LY* ratio obtained by the fit. The slow/total *LY* ratio is plotted versus $\tau_4$ in Fig. 10. If we require the slow/total *LY* ratio to be 0.54±0.05 and 0.59±0.05 in undoped LTB and LTB:$Cu^+$, respectively, as given in the previous section, $\tau_4$ becomes almost the same for undoped LTB and LTB:$Cu^+$, i.e. 21±6 and 20±6 μs , respectively (see Fig. 10). This magnitude of $\tau_4$ is consistent with the results of other groups for the emission peak at 365 nm in LTB:$Cu^+$: 25 μs [18] for excitation at 240 nm and 29 μs [16] for electron beam excitation.

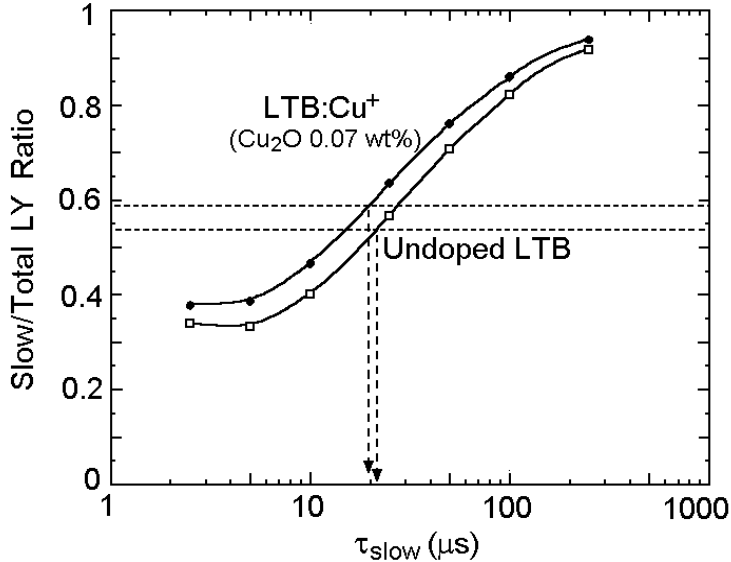

Fig. 10
The slow/total *LY* ratio given by the fit of the decay spectra (Fig. 8) as a function of the decay constant $\tau_4$. Open and solid circles give the fit results. Solid curves are drawn for guiding the eye. The slow/total *LY* ratio of ~0.54 and 0.59 (Table 2) are indicated by dashed lines.

**Radiation damage**

The radiation damage of optical transmission due to $^{60}$Co γ-rays is given in Fig. 11 for a single crystal of LTB:$Cu^+$ ($Cu_2O$ 0.01 wt%). A new absorption peak appeared upon irradiation at around 340 nm (3.65 eV) close to the emission peak, besides old ones at 240 nm (5.17 eV), and 270 nm (4.59 eV) before irradiation. The damage increases with accumulated dose up to $10^4$ Gy and is saturated above it. The saturated value of the induced absorption coefficient at 365 nm is about 26 $m^{-1}$. While there are no data to the authors' knowledge, the radiation damage induced by neutrons is generally expected to be larger than that due to γ-rays because of possible disintegration of nuclei upon collision with neutrons.

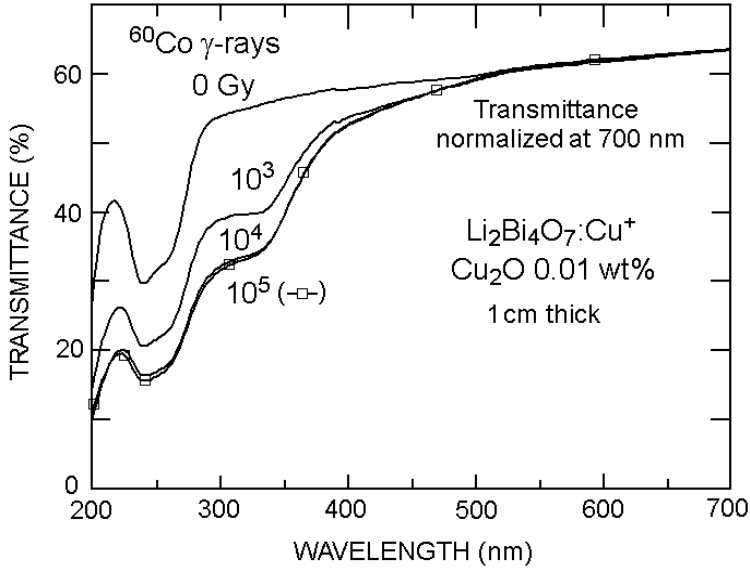

Fig. 11 Radiation damage of a single crystal of LTB:$Cu^+$ ($Cu_2O$ 0.01 wt %) due to $^{60}Co$ γ-rays.

## CONCLUSION AND DISCUSSIONS

The obtained result can be summarized as follows:

**(1)** The time-integrated *LY* for X-rays is about 600 ph/MeV in undoped LTB, and 760 ph/MeV in LTB:$Cu^+$($Cu_2O$ 0.07 wt%), assuming the *LY* of 6000 ph/MeV in 1 cm thick $Bi_4Ge_3O_{12}$ single crystals [24].

**(2)** *LY* for γ-rays within ~1 μs gate was roughly half of the time-integrated *LY* for X-rays in both undoped and Cu-doped LTB.

**(3)** α/γ ratio was 0.15 and 0.18 in undoped and $Cu^+$-doped LTB, respectively. As the $^{241}$Am-α source was external, we cannot exclude a possibility that the α/γ ratio may be degraded to some extent due to the actual crystal surface.

**(4)** The 10-90% rise time was 163 ps for both undoped and $Cu^+$-doped LTB after correction for the *TTS* of the PMT, whose nominal value was 160 ps in FWHM.

**(5)** Decay spectra were measured for 400 and 5000 ns full ranges. Combining the fit for both spectra, we obtained the result given in Table 4. The decay spectra of undoped and $Cu^+$-doped LTB are similar to each other. The slowest component with $τ_4$ = 21±6 μs and 20±6 μs was most dominant in undoped and $Cu^+$-doped LTB, respectively. It is followed by the second dominant component with $τ_3$ = 300-400 ns. The sum of the fastest (with $τ_1$ less than 1 ns) and the next to the fastest ($τ_2$ in a few tens ns) components occupies 7% in undoped LTB and 5% in LTB:$Cu^+$ of the time-integrated *LY*. In Table 4, a finite range is given for each of $τ_2$ and $τ_3$ with its maximum and minimum taken from the fit in 400 ns (Fig. 8) and 5000 ns (Fig. 9) full scales, respectively.

Table 4 Decay characteristics for pulsed X-ray excitation. Decay constants $\tau_1$ - $\tau_4$ are given with the percentage of $LY$ in parentheses.

|  | undoped LTB | LTB:Cu$^+$(Cu$_2$O 0.07 wt%) |
| --- | --- | --- |
| 1-st component | $\tau_1$=0.86 ns (~3%) | $\tau_1$=0.82 ns (~2%) |
| 2-nd component | $\tau_2$=26-59 ns (~4%) | $\tau_2$=29-45 ns (~3%) |
| 3-rd component | $\tau_3$=287-403 ns (~36%) | $\tau_3$=327-385 ns (~31%) |
| 4-th component | $\tau_4^{\#}$=25000 ns (~57%) | $\tau_4^{\#}$=25000 ns (~64%) |

#) $\tau_4$ was fixed at 25 µs in the fit for LTB:Cu$^+$ based on [18]; $\tau_4$ is constrained within 14~26 µs from the present work only. For undoped LTB, the same $\tau_4$ was used according to the present result as shown in Figs. 9-10 (see text).

**(6)** Radiation-induced absorption coefficient $\mu_{ir}$ in LTB:Cu$^+$(Cu$_2$O 0.01 wt%) at 365 nm was 19 m$^{-1}$ at 10$^3$ Gy, and 26 m$^{-1}$ at both 10$^4$ and 10$^5$ Gy for $^{60}$Co γ-rays.

A few discussions are given below:
**(D1)** Increase in $LY$ due to Cu$^+$ doping

The obtained increase of ~30% is not large, while it depends on the Cu$^+$ concentration and crystal quality. This number was obtained for Cu$^+$-doped LTB crystals which give rather high $LY$. The modest increase in $LY$ may be due to the following reasons: **(i)** energy transfer from the host to Cu$^+$ is not significant [18]; **(ii)** host emission is partly absorbed in the absorption band of Cu$^+$. **(iii)** Cu$^+$ concentration is limited so as to avoid segregation [9].

**(D2)** $LY$ for thermal neutrons n$_{th}$:

n$_{th}$ capture in LTB is dominated by $^{10}$B+n$_{th}$→α(1.47 MeV)+$^7$Li*(0.84)+ γ (0.48) with a branching ratio of 94%. Contribution of $^7$Li+ n$_{th}$ is small because both natural abundance and n$_{th}$ capture cross section of $^7$Li (7.5% and 940 b, respectively) are much smaller than for $^{10}$B (20% and 3830 b). The 0.48 MeV γ-ray is assumed to leave before absorption. As the γ-ray equivalent energy $GEE$ is known for neither 1.47 MeV α nor 0.84 MeV $^7$Li*, we assume the ratio ξ of $GEE$(n$_{th}$)/$GEE$(5.5 MeV α) in LTB to be the same as in B-loaded PS (Table 1). ξ in B-loaded PS becomes 0.1/0.55=0.18 by using $GEE$ (n$_{th}$)~0.1 MeV [26,27] and $GEE$(5.5 MeV α)~0.55 MeV which is usually taken for plastic scintillators. Then, $GEE$(n$_{th}$) in LTB is given by 0.18x$GEE$(5.5 MeV α)=0.18x0.18x5.5 MeV ~ 0.18 MeV by using the α/γ ratio of 0.18 at 5.5 MeV obtained in the present work. Then n$_{th}$ capture in LTB:Cu$^+$ gives $LY$(n$_{th}$)=$GEE$(n$_{th}$)x311 ph/MeV (Table 2) ~56 ph/n$_{th}$ within 1 µs. To increase $LY$(n$_{th}$), we also developed Ce$^{3+}$-doped Li-B-O glass [23,28, 29]. $LY$(n$_{th}$) is increased 4-5 times and the slow component disappears. While increase in the B concentration increases *LY* further, it causes stronger hygroscopicity.

**(D3)** Decay kinetics

The obtained result can be compared with past works. For undoped LTB, fast emission due to SOR excitation consisted of $\tau$=10 and 150 ns components with the *LY* ratio of 19:81 [14]. Besides them, a dominant slow component with $\tau$>10 µs was claimed [15,20]. For LTB:$Cu^+$, there are no measurements published for fast ($\tau$<1 µs) components in the past to the authors' knowledge. For the slow one, $\tau$ was 25 µs for the 365 nm emission peak excited at ~240 nm by a 1 µs flush tube [18], and 29 µs in radioluminescence excited by an electron beam of 2.6 µs duration [16].

The present work shows that the decay kinetics of radioluminescence in both undoped and $Cu^+$-doped LTB are similar to each other, when measured not at a specific wavelength of 365 nm but for the whole emission peak. This result suggests that $Cu^+$-induced emission has both fast and slow components. To suppress the host emission, an additional measurement was carried out by inserting a low-cutting filter UV35 (transmitting above 350 nm) between the crystal and PMT. The obtained decay spectrum was again similar to undoped LTB as well as LTB:$Cu^+$ without the filter. However it is not yet conclusive from this result whether $Cu^+$-induced emission contains both slow and fast components, as the fast component due to the host may remain above 350 nm. There is, however, another evidence. If $Cu^+$-induced emission is totally slow, sizable increase in *LY* observed for $\alpha$- and $\gamma$-rays within 1 µs due to $Cu^+$ doping (Eqs. 2-3) could not be explained.

**(D4)** $Cu^+$ emission mechanism

According to [18], it is natural to take the emission peak at 365 nm in LTB:$Cu^+$ due to the parity forbidden slow $3d^94s^1 \rightarrow 3d^{10}$ radiative transition of $Cu^+$. However, there remain some problems to be studied further:

**(i)** Besides the slow component with $\tau$=25 µs [18], there are also sizable fast components in the $Cu^+$-induced emission as mentioned above in (D3). What is the emission mechanism of this fast component?

**(ii)** Undoped LTB has a dominant slow decay in a range of $\tau$=15-27 µs (see "Decay kinetics" in "RESULTS"). $\tau$ of 25 µs [18] for the 365 nm emission peak assigned to $Cu^+$ intracenter transition is in the same range as above. Is this accidental or related to the mechanism of $Cu^+$-induced emission?

Acknowledgements

This research was supported by the Ministry of Education, Culture, Sports, Science and Technology of Japan under Grant-in-Aid for Scientific Research.

# References

[1] C.W.E. van Eijk, *Proc. 5th Int. Conf. on Inorganic Scintillators and their Use in Scientific and Industrial Applications (SCINT99)*, Moscow, 1999, p.22.
[2] Bicron Co. *Brochure*, 1990.
[3] A.R. Spowart, Nucl. Instr. Meth. 135(1976)441.
[4] A.R. Spowart, Nucl. Instr. Meth. 140(1977)19.
[5] J.B. Birks, *The Theory and Practice of Scintillation Counting*, (Pergamon, Oxford, 1964), p.544.
[6] C.W.E. van Eijk, Radiat. Prot. Dosim. 110(2004) 5.
[7] C.M. Combes, *Scintillation properties of $^6$Li-based materials for thermal -neutron detection*, Ph.D thesis, Delft University of Technology, 1999.
[8] A. Gektin et al., *Proc. 4th Int. Conf. on Inorganic Scintillators and their Use in Scientific and Industrial Applications (SCINT97)*, Shanghai, 1997, p.121.
[9] M. Ishii et al., J. Cryst. Growth 257(2003)169.
[10] Landolt-Börnstein, *New Series Group III*, *Crystal and solid state physics*, Vol. 7, *Part d2*, (Springer, Berlin, 1980).
[11] M.M. Islam et al., J. Phys. Chem. 109(2005)13597.
[12] A.Yu. Kuznetsov et al., Phys. Solid. State.41(1999)48.
[13] V.T. Adamiv et al., Materials 3(2010)4550.
[14] I.N. Ogorodnikov et al., Nucl. Instr. Meth. A448(2000)467.
[15] I.N. Ogorodnikov et al., Phys. Solid State 42(2000)464.
[16] M. Ignatovych et al., Rad. Phys. Chem. 67(2003)587.
[17] M. Ignatovych et al., Radiat. Meas. 38(2004)567.
[18] G. Corradi et al., J. Phys.: Condens. Matter 20(2008)025216.
[19] N. Senguttuvan et al., Nucl. Instr., Meth. A486(2002)264.
[20] V.V. Chernikov et al., Nucl. Instr. Meth. A498(2003)424.
[21] D.M. Knotter et al., Inorg. Chem. 31(1992)2196.
[22] M. Kobayashi et al., Nucl. Instr. Meth. A592(2008)369.
[23] M. Ishii et al., Radiat. Meas. 38(2004)571.
[24] M. Moszynski et al., IEEE Trnas. Nucl. Sci. NS-44(1997)1052.
[25] N. Kudomi, Nucl. Instr. Meth. 430(1999)96.
[26] M.C. Miller et al., Nucl. Instr. Meth. A422(1999)89.
[27] G.I. Britvich et al., Nucl. Instr. Meth. A550(2005)343.
[28] M. Ishii et al., Nucl. Instr. Meth. A537(2005)282.
[29] M. Ishii et al., *Proc. KEK-RCNP Int. School and Mini-WS for Scintillating Crystals and Their Applications in Particle and Nuclear Physics*, Tsukuba, 2003, KEK Proceedings 2004-4, p.99.